\documentclass[prd,twocolumn,floatfix]{revtex4}
\usepackage[dvips]{graphics}
\usepackage{amsmath,amsfonts}
\usepackage{psfig}

\begin{document}

\newcommand{\bra}{\langle}
\newcommand{\ket}{\rangle}
\newcommand{\beq}{\begin{equation}}
\newcommand{\eeq}{\end{equation}}
\newcommand{\be}{\begin{equation}}
\newcommand{\ee}{\end{equation}}
\newcommand{\bea}{\begin{eqnarray}}
\newcommand{\eea}{\end{eqnarray}}
\def\bea{\begin{eqnarray}}
\def\eea{\end{eqnarray}}
\def\tr{{\rm tr}\,}
\def\href#1#2{#2}

\preprint{UW/PT-02/25}

\title{Hadron Masses and Screening from AdS Wilson Loops}

\author{Andreas Karch} \email{karch@phys.washington.edu}
\author{Emanuel Katz} \email{amikatz@phys.washington.edu}
\author{Neal Weiner} \email{nealw@phys.washington.edu}
\affiliation{Department of Physics,
University of Washington,
Seattle, WA 98195, USA}

\begin{abstract}
We show that in strongly coupled ${\cal N} =4$ SYM the binding
energy of a heavy and a light quark is independent of
the strength of the coupling constant. As a consequence
we are able to show that in the presence of light
quarks the analog of the QCD string can snap and color
charges are screened. The resulting neutral mesons interact
with each other only via pion exchange and we estimate the masses
of those states.
\end{abstract}
\today
\maketitle

\section{Introduction}

\begin{figure}
 \centerline{\psfig{figure=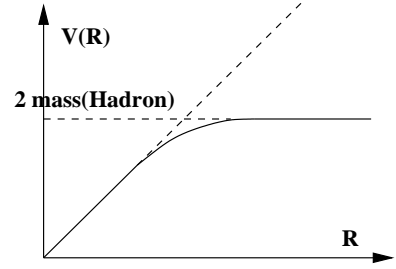,width=2.0in}}
 \caption{
Quark-antiquark potential in a confining gauge theory with
dynamical fundamental matter. After an initial linear rise
the potential plateaus at twice the mass of the lightest hadron.
}
 \label{linear}
  \end{figure}

In QCD probably the
most dramatic modification of the dynamics due to the presence
of dynamical flavors is the string breaking: as one tries to
separate two infinitely heavy test quarks (whose physics can
be neatly captured by Wilson loops) the potential does not rise
linearly with the distance forever, as it should in a
confining theory. Instead, once the energy stored in the flux tube
is big enough, the string snaps by pair producing a dynamical
quark-antiquark pair. The interactions between the two resulting mesons 
rapidly die out with distance. The
potential, as probed by the Wilson line, rises linearly for a while and
then saturates at twice the mass of the meson, see Fig.\ref{linear}.
One can consider the ``probe limit" of finite number of flavors, $M$, 
and large number of colors, $N$, 
with $g^2 N$ fixed and hence $g^2 M \rightarrow 0$. Since
the actual snapping process has an amplitude proportional
to $g^2 M$,
in the probe limit the transition becomes sharp. The potential is completely dominated by the lowest energy configuration.

In \cite{Maldacena:1998re} 
it was shown that ${\cal N}=4$ super Yang-Mills conformal field
theory is dual to type IIB 
string theory on a background of $AdS_5 \times S_5$. 
Dynamical flavors with a finite mass can be included in the
supergravity description of strongly coupled gauge theories via
spacetime filling D-branes \cite{Karch:2002sh}, at least in the probe limit.
In this letter we will investigate to what extent the
prescription of \cite{Karch:2002sh} can be used to study
the behavior of the QCD string in the presence of dynamical quarks. 
In particular, we will find additional 
configurations where strings stretch between the boundary and the D7 branes, yielding a simple interpretation of string snapping in the string dual description.

\begin{figure}
 \centerline{\psfig{figure=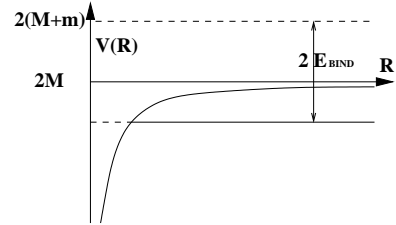,width=2.0in}}
 \caption{
Quark-antiquark potential between two heavy test quarks of mass
$M$ in a conformal field theory in the presence
of a dynamical quark of mass $m$. At weak coupling the binding energy
is less than the mass of the dynamical quark and hence the potential
is Coulomb all the way. If the binding energy becomes bigger than $m$
one finds a crossover from Coulomb behavior to a constant.
}
 \label{coulomb}
  \end{figure}

For light quarks QCD is never truly confining in the sense of
linearly rising potentials due to flux strings. However the
main experimental fact is still true: colored quarks can never
appear in isolation. They always pair produce a light quark-antiquark
pair to form color neutral mesons. Note that latter can
in principle also happen in a conformal theory. In a conformal
theory the potential is forced to be Coulomb by dimensional
analysis. So there is only a finite cost to separate two heavy quarks of mass
$M$, and the total energy of the configuration asymptotes to $2M$ for
large separation. However if the mass of the produced
meson
$$m_{Meson} = M +m - E_{Bind}$$
is less than $M$, it is still energetically favorable for the string
to snap, see Fig.\ref{coulomb}. 

Note that for weakly coupled Coulomb potentials this never happens.
There the strength of the Coulomb potential is proportional
to the coupling constant, that is e.g. $g^2 N$ for ${\cal N}=4$ SYM,
and hence the binding energy goes like $E_{Bind} \sim (g^2 N)^2 m $
which at weak coupling is always less than $m$. For example for QED
the binding energy for the lightest meson (the hydrogen atom)
formed out of the ``heavy
test quark" (the proton) and the ``light dynamical quark" (the electron)
is 13.6 eV, which is roughly a factor of $\frac{1}{137^2}$ down
from 0.5 MeV, the mass of the dynamical quark. So protons do not pop
electrons out of the vacuum to shield their charge. However
for strong coupling it seems that the binding energy can easily
be as big or bigger than $m$. Consequently, in a strongly coupled CFT the 
``Coulomb string" can snap, just like the ``Confining string".
As in QCD, the charged heavy quarks can never
be studied in isolation, since they'll always pair produce dynamical quarks
to form neutral mesons, color is screened.

In this letter we will use the AdS/CFT correspondence to 
show that that in strongly coupled
${\cal N}=4$ SYM the binding energy at large $g^2 N$ actually
becomes independent of $g^2 N$, it is an order one number
times the mass of the light quark. The order one constant depends
on the R-charge orientation of the quark. So in ${\cal N}=4$
SYM the string can snap and the procedure of \cite{Karch:2002sh} is
able to give a quantitative description of this effect. We expect
the snapping of the string in a real confining theory to be
described in precisely the same fashion.

\section{Supergravity Evaluation of the Wilson Loops}
It was proposed in \cite{Maldacena:1998im,rey} 
that Wilson loops of the CFT can be described in AdS by
\be
\langle W ({\mathcal C}) \rangle \sim \lim_{\Phi \rightarrow \infty} e^{-(S_\Phi - l \Phi)},
\ee
where $S$ is the proper area of a fundamental string worldsheet which lies on 
the loop $\mathcal C$ on the boundary of AdS, $l$ is the total length of the 
Wilson loop and $\Phi$ is the mass of the W boson. For us, the appropriate quantity will be a suitable Legendre transform \cite{Drukker:1999zq}, 
as we will discuss shortly. The heavy quark is 
constructed by taking one brane and sending it to infinity. As 
this brane goes to infinity, its dynamics decouple and strings stretching 
between it and the boundary act as infinitely massive sources of color charge. 

\begin{figure}
 \centerline{\psfig{figure=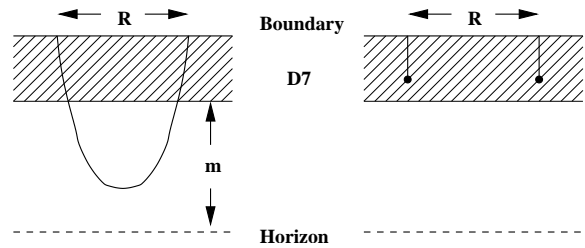,width=3.0in}}
 \caption{The two classical configurations contributing to
the expectation value of the Wilson line in the presence of
dynamical quarks.} \label{fig:configs}
  \end{figure}

The energy of the first configuration shown in Fig.\ref{fig:configs} was
calculated in \cite{Maldacena:1998im} to be
\be
E=-\frac{2 (2 g_{YM}^2 N)^{1/2}}{\pi L}(1-l^2(\theta_i,\phi_i))^{3/2} I_1^2 (l(\theta_i,\phi_i)),
\label{eq:malden}
\ee
where
\bea
I_1(l) = \frac{1}{(1-l^2) \sqrt{2-l^2}}\times \\
\nonumber \left[(2-l^2) 
E\left(\frac{\pi}{2},\sqrt{\frac{1-l^2}{2-l^2}}\right)-
F\left(\frac{\pi}{2},\sqrt{\frac{1-l^2}{2-l^2}}\right)
\right],
\eea
and $L$ is the separation between the heavy quarks, and $F,\ E$ are elliptic integrals of the first and second kind, respectively. The quantity $l$ is defined
by in terms of the angular separation of the two heavy quarks by
\be
\frac{\Delta \theta}{2} = \frac{l}{\sqrt{2-l^2}} F\left(\frac{\pi}{2},\sqrt{\frac{1-l^2}{2-l^2}}\right).
\ee

In the presence of dynamical light quarks, described by the 
presence of the D7 branes, there is
a second set of consistent boundary conditions. The Wilson loop will
be given by the sum of the exponentials of the actions associated with each configuration. 
\beq
 <W ( {\cal C})> = \sum_i e^{-\bar S_i},
\eeq
where $\bar S$ is the suitably regulated action.
For distances $L \ll m^{-1}$ we expect the first configuration
to dominate the Wilson loop, while for $L \gg m^{-1}$ we expect the meson pair
configuration to dominate.

Actually the action of the second configuration can be
assembled out of three contributions. The energy of the two string pieces
and the energy of the flux that connects the two string endpoints.
In the field theory these three contributions correspond to the mass
of the two heavy-light mesons and the potential energy between them.

\subsection{Heavy-Light Binding Energy}

The string will be fixed in the 3-dimensional space coordinates as we vary 
$\sigma$, and will lie along a great circle, so that its position 
in $S_5$ can be parameterized by one coordinate $\theta$. The metric is then
\be
ds^2 = \alpha' \left( \frac{U^2}{R^2} dt^2 + \frac{R^2}{U^2} dU^2 + R^2 d \Theta^2 \right).
\ee
We want time independent solutions, so we set $\tau=t$ 
and since the string will not double-back in $U$, we can set $\sigma = U$. The 
induced metric on the worldsheet is then
\be
\alpha' \ \begin{pmatrix}
\frac{U^2}{R^2} & 0 \cr
0 & \frac{R^2}{U^2} +\left(\frac{\partial \Theta}{\partial U} \right)^2 R^2 
\end{pmatrix}
\ee
The resulting action is 
\be
S = \frac{T}{2 \pi} \int dU \ \left( 1+ U^2 \left( \frac{\partial \Theta}{\partial U} \right)^2 \right)^{1/2}.
\label{eq:action}
\ee
Notice that this action is completely independent of R.

Interestingly, this is just the metric for a string in flat space! It is convenient to change coordinates to those where the D7 brane lies along the one-parameter family $(0,2 \pi m,z)$ where $m$ specifies the mass of the quarks.

We parameterize the position of our heavy quark brane by $(x_0, y_0, z_0)$, which we will eventually take to infinity. In this case, the path from the heavy quark brane to the light quark brane will be given by
\be
(x(t),y(t),z(t)) = (x_0 , y_0-2\pi m,0)t + (0,2 \pi m,z_0)
\ee

The appropriate quantity to calculate is not the area of the worldsheet, but
a Legendre transform of it \cite{Drukker:1999zq}. The energy is
\be
E=\frac{S}{T}=\frac{1}{2 \pi} \tilde A = \frac{1}{2 \pi}(A - P_i Y^i),
\ee
where
\be
P_i = \frac{\delta A}{\delta \partial_\sigma Y^i}.
\ee
Setting $\sigma=t$ we find
\bea
\tilde A &=& \sqrt{x_0^2 + (y_0-2 \pi m)^2}-\frac{x_0^2 + y_0(y_0-2 \pi m)}{\sqrt{x_0^2 + (y_0-2 \pi m)^2}} \\ \nonumber
&=& \frac{2 \pi m (2 \pi m-y_0)}{\sqrt{x_0^2 + (y_0-2 \pi m)^2}}.
\eea
We parameterize the position of the distant brane by
\be
(x_0,y_0,z_0) = M (\sin \theta \sin \phi, \sin \theta \cos \phi, \cos \theta).
\ee
We will see shortly that the parameter $\phi$ controls the strength of the heavy quark-light quark interaction. In the $M\rightarrow \infty$ limit we find
\be
\tilde A= -m \cos \phi.
\label{eq:d7en}
\ee
The binding energy of the system is
\be
E_{bind}=2m( \cos^2(\phi_1/2)+\cos^2(\phi_2/2))
\ee
For order one angles, the screening length (beyond which the meson configuration will dominate), is given by
\be
L_{sc} \sim \frac{2 \sqrt{g^2_{YM} N}}{m}.
\ee

For comparison one can also study the case of a probe D3 brane
instead of the probe D7 brane. This corresponds to adding
dynamical W-bosons instead of the dynamical hypermultiplets:
the dual gauge theory is ${\cal N}=4$ $SU(N +M)$ broken down to $SU(N) \times
U(M)$. For $M<<N$ the probe approximation can be used,
even though in this case it is not necessary since
the full metric is known.
In the probe limit
a very similar calculation shows that in this case the binding
energy is independent of $g^2 N$ at large $g^2 N$ as well and just given
by $\cos(\alpha) m$, where $\alpha$ is the angle between the heavy
quark and the dynamical W-boson on the $S^5$.
The action we have to minimize
is still given by (\ref{eq:action}), so again we are interested
in the geometry of straight lines in flat 6d space.
The stack of $N$ D3 branes is sitting at the origin. We choose
our coordinates such that the probe flavor brane is sitting
at 
$x= 2 \pi m$ 
with all other coordinates being zero, and
the distant brane at
$(x,y) = M (\cos(\alpha),\sin(\alpha))$.
That is there are 3 special points in $R^6$ and we pick $x$,$y$ to
be the cartesian coordinates of the plane defined by those 3 points.
The corresponding area is
\bea
\nonumber
\tilde{A} &=& \frac{\sqrt{ (M \cos(\alpha) -2\pi m)^2 + M^2 \sin^2 (\alpha)} - M}{2\pi}=\\
&=& - \cos(\alpha) m
\eea
and hence the binding energy becomes 
\be
E=2m (\cos^2(\alpha_1/2)+\cos^2(\alpha_2/2)).
\ee
Even though the bound
state energy is comparable in the two cases, the physics in the end
is quite different, since the meson-meson interactions
differ significantly.

\subsection{Meson - Meson potential}

\begin{figure}
 \centerline{\psfig{figure=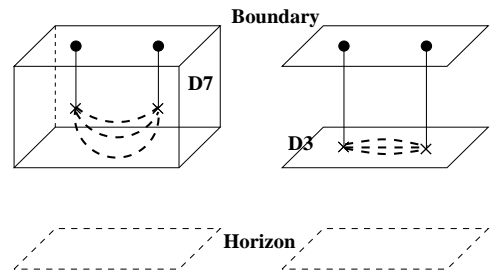,width=2.5in}}
 \caption{
The flux dual to the meson-meson potential for a dynamical HM from
D7 branes and a dynamical W-boson from D3 branes respectively.
In latter case the flux is confined to a plane and bound to give rise
to a Coulomb force. In the former the flux can make use
of the AdS geometry and leads to no long range force.}
 \label{mesmes}
\end{figure}

In order to capture the contribution of the meson-meson potential
to the Wilson loop, we have to calculate the contribution
to the action due to the flux lines connecting the ends of the open
strings. Fig.\ref{mesmes} displays the flux for the two cases
of D7 brane and D3 brane.

In the case of the D3 brane, the flux lines are confined to a Poincare
slice and just behave like flux lines in flat space. They give rise to
a Coulomb interaction, due to the exchange of the massless
gauge fields living on the D3 brane. What happens in the dual field theory
is that our mesons are charged under the $U(M)$ gauge fields in 
$SU(N+M) \rightarrow SU(N) \times U(M)$. The dynamical W-bosons
after all transform in the bifundamental representation of $SU(N) \times
SU(M)$. The $SU(N)$ charge of the heavy quarks gets screened as before
once the binding energy becomes bigger than the mass of the W,
that is for
appropriate choices of $\alpha$, the relative R-charge orientation.
Instead of the crossover from Coulomb to constant behavior, we get
a crossover from the strong Coulomb of the $SU(N)$ gauge group
(with order one potentials independent of $g^2 N$) to constant plus
weak Coulomb (with potentials of order $g^2 M$) due to the massless $U(M)$
gauge boson exchange.

The situation becomes more interesting in the case of the D7 branes.
Here the interactions between the heavy-light mesons are due to
pion exchange, that is exchange of light-light mesons. The
pions are massive and hence the interaction is exponentially
suppressed. The masses of the pions are the masses of the KK-modes
of the worldvolume gauge field on the D7 brane. For zero mass quarks,
the D7 brane is AdS$_5$ filling and one has a continuum of states.
For finite mass quarks the D7 brane ends and the KK-spectrum
becomes discrete with a mass gap. This is analogous to what
happens to the graviton in a confining gauge theory: The continuous
spectrum of the AdS$_5$ background becomes a discrete
spectrum with a mass gap, the modes corresponding to the glueballs
of the gauge theory in that case.
To calculate the energy in the flux explicitly, one calculates the
Green's function for the gauge field on the D-brane. This can be done
in a mode expansion, which makes it obvious that also from the
bulk perspective the meson-meson potential is dominated by
the lightest KK-mode of the gauge field.

In order to find the mass of the lightest KK-mode we would have to solve
Maxwell's equation on the curved D7-brane worldvolume. In order
to at least determine the coupling constant dependence we can consider
a somewhat simpler toy model: instead of the position
dependent tension of the D7 brane, we approximate it 
with a simple box shape:
\be
T(U) \propto \left( \sqrt{1-(2 \pi m)^2 / U^2}\right )^3 \rightarrow
\left \{ \begin{array}{ll} 
1& \; \; \; \mbox{ for } U> 2 \pi m \\ 0& \; \; \; \mbox{ for } U\leq 
2 \pi m
\end{array} \right .
\ee
This is merely a study of gauge fields in AdS with an IR cutoff which has been 
studied already \cite{Davoudiasl:1999tf,Pomarol:1999ad,Kaloper:2000xa}.
Regularity at the UV boundary requires that only the $J_1$ Bessel function
contributes. 
We can consider states with either Neumann and Dirichlet boundary conditions
on this IR brane which then demands
that mass of the $n$-th mode satisfies:
\be
\left . J_1(m_n/U) \right |_{U=2 \pi m} =0,
\ee
or
\be
\left . J_1'(m_n/U) \right|_{U= 2 \pi m} = 0.
\ee
The masses of the pions are order one numbers times the mass of the quarks.
This is not surprising: just like the heavy-light binding energy
the light-light binding energy is an order 1 number times the mass
of the quarks $m$. Like the heavy-light binding energy it never
becomes equal or bigger than $2m$, which would correspond to massless
or tachyonic pions.

\vskip10pt

\section{Conclusions}
One can exploit the AdS/CFT correspondence to gain great insight into 
questions related to the qualitative behavior of QCD. We have seen that
the inclusion of dynamical flavors into N=4 SYM gives a simple interpretation
of QCD string snapping in the gravity dual. 
Although there is no confinement in these sense of 
area law Wilson loops, all asymptotic states are color neutral and color 
charge is completely screened. This furthers the hope that many more
interesting features of QCD can be understood using string duals.

\vskip15pt

\noindent{\large\bf Acknowledgments:}
We would like to thank Hirosi Ooguri, Matt Strassler and Juan Maldacena
for helpful comments.
This work was partially supported
by the DOE under contract DE-FGO3-96-ER40956.

\bibliography{confine}
\bibliographystyle{apsrev}
\end{document}